# Raman spectroscopy and strain mapping in individual Ge-Si$_x$Ge$_{1-x}$ core-shell nanowires

D. C. Dillen, K. M. Varahramyan, C. M. Corbet, and E. Tutuc

*Microelectronics Research Center, University of Texas, Austin, Texas 78758*

Core-shell Ge-Si$_x$Ge$_{1-x}$ nanowires (NWs) are expected to contain large strain fields due to the lattice-mismatch at the core/shell interface. Here we report the measurement of core strain in a NW heterostructure using Raman spectroscopy. We compare the Raman spectra, and the frequency of the Ge-Ge mode measured in individual Ge-Si$_{0.5}$Ge$_{0.5}$ core-shell, and bare Ge NWs. We find that the Ge-Ge mode frequency is diameter-independent in GeNWs with a value similar to that of bulk Ge, 300.5 cm$^{-1}$. On the other hand, Ge-Si$_{0.5}$Ge$_{0.5}$ core-shell nanowires reveal a strain-induced blue shift of the Ge-Ge mode, dependent on the relative core and shell thicknesses. Using lattice dynamical theory we determine the strain in the Ge core, and show that the results are in good agreement with values calculated using a continuum elasticity model.

## I. INTRODUCTION

Semiconductor nanowires (NWs) have attracted considerable attention recently both as a novel quasi one-dimensional platform to study electron physics[1-3], and as a possible replacement for the channel material of a planar metal-oxide-semiconductor field-effect transistor[4-6] (MOSFET) thanks to superior scaling properties in the gate-all-around geometry.[7] The carrier mobility may also be enhanced in these materials through the use of band engineered radial heterostructures.[8-10] More recently, the existence of helical states which lock the hole spin and wave-vector, has been theoretically predicted in Ge-Si core-shell nanowires.[11] Furthermore, group IV core-shell nanowire heterostructures maintain compatibility with traditional complementary metal-oxide-semiconductor technology.

The use of radial epitaxial heterostructures, especially in the Ge-Si materials system, is expected to produce large elastic strain fields due to the mismatch of equilibrium lattice spacing at the core-shell interface. This strain can change the energy momentum dispersion and energy band offset in the heterostructure, and consequently the effective mass and carrier mobility. Therefore, probing the elastic strain in such heterostructures is beneficial to the design and growth of core-shell nanowires, as well as to the understanding of their electrical properties. The strain distribution of core-shell nanowires, which is markedly different than that of planar heterostructures, has been studied theoretically by several groups.[12-15] Here we present an experimental study of the strain in individual Ge-Si$_x$Ge$_{1-x}$ core-shell nanowires, determined using Raman spectroscopy combined with lattice dynamic theory.

## II. EXPERIMENTAL

The Ge-Si$_x$Ge$_{1-x}$ core-shell NWs investigated here were grown on a Si (111) wafer by a Au-catalyzed vapor-liquid-solid (VLS) mechanism, followed by an epitaxial shell growth using ultra-high-vacuum chemical vapor deposition (UHVCVD). A schematic representation of core-shell nanowire growth is shown in Fig. 1(a). By varying the relative SiH$_4$ and GeH$_4$ gas flow rate during shell growth, Ge-Si$_x$Ge$_{1-x}$ core-shell NWs can be grown with controllable shell compositions.[16] In this study we use core-shell NWs with a shell Si content of $x = 0.5$, and thickness $t_{sh} = 5$ nm, as determined by transmission electron microscopy (TEM) and energy dispersive X-ray (EDX) spectroscopy. The finite, albeit small, conformal Ge growth during VLS core growth results in a tapering of the core's diameter from the base to the tip. The shell, however, is expected to have a constant thickness along the NW length. We also grow bare Ge NWs as a baseline comparison; their growth is identical to that of the Ge core in the core-shell NWs described above. Figure 1(b) shows a TEM micrograph of a Ge-$Si_{0.5}Ge_{0.5}$ NW, evincing the single crystal structure, and the epitaxial $Si_{0.5}Ge_{0.5}$ shell growth. Furthermore, the NWs appear cylindrical in shape, with no discernible faceting observed in cross-sectional TEM imaging.

Post growth, the samples were prepared for Raman characterization by sonication in ethanol, followed by drop-casting onto patterned glass substrates. The glass substrate used here was chosen to eliminate the overlap of the Raman signal originating from the nanowire with that of the substrate.[17]

We have obtained the Raman spectra from individual nanowires using a Renishaw InVia $\mu$-Raman Spectrometer in backscattering geometry, with a polarized 532 nm excitation source, focused to a spot of ~1 $\mu m$ in diameter using a 100x objective lens. An incident laser power of 13 kW/cm$^2$ was used throughout, a value sufficiently small to eliminate red-shifting of the Ge-Ge Raman peak due to sample heating.[18] No polarizer was used in the collected signal path.

In order to examine the diameter dependence of Raman spectra, we collected data at multiple points along the axis of individual nanowires. The nanowire diameter ($d$) at each measurement position was determined by atomic force microscopy (AFM). The $d$-values range from 15 to 50 nm in GeNWs, and from 40 to 65 nm in core-shell samples. Figure 2 shows a comparison between typical Raman spectra, collected from an individual GeNW ($d$=25 nm) and a Ge-$Si_{0.5}Ge_{0.5}$ ($d$=44 nm) core-shell nanowire. The GeNW spectrum shows a single peak with a Raman frequency of 300.5 cm$^{-1}$, a value nearly identical to that of the Ge-Ge mode in bulk Ge, indicating the GeNW is unstrained. On the other hand, the Ge-Ge mode of the core-shell sample is shifted to 305.9 cm$^{-1}$, where the shift direction is consistent with compressive uniaxially strained Ge.[19] We also observe additional Raman peaks near 400 cm$^{-1}$ and 500 cm$^{-1}$ in Ge-$Si_{0.5}Ge_{0.5}$ core-shell nanowires, with an order of magnitude lower intensity, which we attribute to the $Si_xGe_{1-x}$ shell's local Si-Ge and Si-Si modes, respectively.[20–24] Based on the relative intensity of the peak at ~300 cm$^{-1}$ in comparison to the peaks located at 400 cm$^{-1}$ and 500 cm$^{-1}$, we conclude the Ge-Ge peak observed at ~300 cm$^{-1}$ originates from the Ge core, a finding consistent with the relative volume ratio of the two regions. In both GeNWs and Ge-Si$_x$Ge$_{1-x}$ core-shell NWs, we have observed only a single peak at ~300 cm$^{-1}$. In each measurement, we have found the exact Ge-Ge Raman frequency by fitting the data to a Voigt profile between 200 cm$^{-1}$ and 400 cm$^{-1}$, using both the Gaussian and Lorentzian linewidths as fitting parameters.

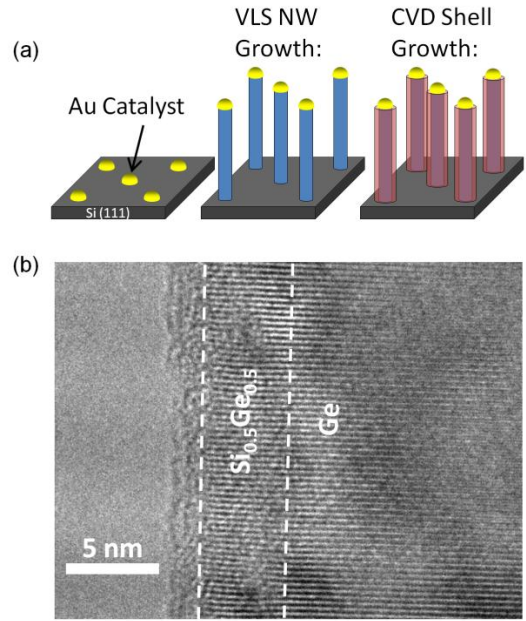

FIG. 1. (Color online) (a) Schematic of Ge-$Si_xGe_{1-x}$ core-shell nanowire growth showing Au catalyst droplets, VLS Ge core growth, and UHVCVD SiGe shell growth. (b) TEM image of Ge-$Si_{0.5}Ge_{0.5}$ core-shell nanowire where the interface has been marked for clarity.

Figure 2 (inset) shows the Ge-Ge Raman peak intensity from a GeNW as a function of the incident polarization angle, defined with respect to the nanowire axis. When the incident electric field is aligned parallel to the nanowire axis, we observe a strong Raman signal. Rotating the incident field polarization away from the nanowire axis causes a decline in the measured peak height, which eventually becomes nearly extinct at perpendicular alignment. The $\cos^2(\theta)$ dependence of the Raman intensity in Fig. 2(inset) indicates the nanowire is acting as an anisotropic antenna.[25–27] Indeed, electrostatic considerations indicate that for a freestanding GeNW in air, the ratio of the local electric field magnitudes for polarizations parallel and perpendicular to its main axis is 8.6.[26] In all subsequent measurements, we align the incident polarization to within 45° of the nanowire axis. Thanks to this parallel electric field enhancement,[26] the incident radiation with electric field polarized along the [111] crystal direction will be transmitted into the nanowire with an approximately order of magnitude greater intensity than the component polarized perpendicular to the nanowire axis.

### III. RESULTS AND DISCUSSION

To provide a baseline for the strain-induced shift of the Ge-Ge vibrational mode in core-shell nanowires, we first performed Raman measurements on unstrained GeNWs of various diameters. We find that the unstrained Ge-Ge mode frequency is constant as a function of nanowire diameter, with an average value of 300.5 cm$^{-1}$. The lack of diameter dependence on the Raman spectra of GeNWs indicates that the diameter of our nanowires is sufficiently large to neglect peak shifts due to phonon confinement and relaxation of the $q\approx 0$ selection rule.[28,29] Subsequent Raman measurements of the Ge-Ge peak position in strained core-shell samples will not be corrected for such factors.

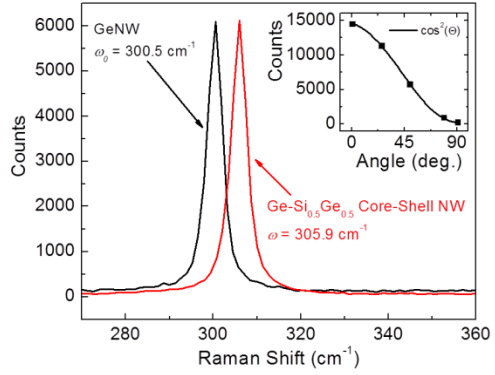

FIG. 2. (Color online) Comparison of the Raman spectra measured in a GeNW (black line) and a Ge-Si$_{0.5}$Ge$_{0.5}$ core-shell (red line) nanowire. The strained Ge-Ge Raman mode of the core is shifted to 305.9 cm$^{-1}$ from the unstrained value of 300.5 cm$^{-1}$. Inset: Intensity of the 300.5 cm$^{-1}$ peak measured in a GeNW as a function of the incident polarization angle.

In the presence of strain, the longitudinal optical (LO) and transverse optical (TO) phonon modes of a cubic crystal, initially degenerate at zone center, will undergo shifts to lower or higher frequencies, depending on the magnitude and sign of the strain (i.e. tensile or compressive). This strain induced frequency shift can be quantitatively described by the secular equation of lattice dynamical theory:[30,31]

$$\begin{vmatrix} p\varepsilon'_{11} + q(\varepsilon'_{22}+\varepsilon'_{33}) - \lambda_1 & 2r\varepsilon'_{12} & 2r\varepsilon'_{13} \\ 2r\varepsilon'_{12} & p\varepsilon'_{22} + q(\varepsilon'_{11}+\varepsilon'_{33}) - \lambda_2 & 2r\varepsilon'_{23} \\ 2r\varepsilon'_{13} & 2r\varepsilon'_{23} & p\varepsilon'_{33} + q(\varepsilon'_{11}+\varepsilon'_{22}) - \lambda_3 \end{vmatrix} = 0. \quad (1)$$

The terms $p$, $q$, and $r$ are the Ge core phonon deformation potential (PDP) values, which relate the measured, by Raman spectroscopy in this case, optical phonon frequencies ($\omega_i$), and the applied strain tensor, $\varepsilon'$. The PDP values have been determined experimentally for bulk Ge by polarized Raman spectroscopy as a function of an applied stress along different crystal directions.[19,32] The strain induced shift of the Raman peak is described by $\lambda_i = \omega_i^2 - \omega_{i0}^2$ where $\omega_{i0} = 300.5$ cm$^{-1}$ is the unstrained frequency of the Ge-Ge mode.

The secular equation can be simplified by considering the form of the core's strain tensor expected when cylindrical symmetry is assumed.[12–14] We employ a NW-oriented Cartesian coordinate system with the $z$-axis along the [111] crystal direction and nanowire main axis, and with the $x$- and $y$-axis defined by the [1$\bar{1}$0] and [11$\bar{2}$] crystal directions, respectively. In this coordinate system, the core's strain tensor is expected to take the form:[12–14]

$$\boldsymbol{\varepsilon} = \begin{pmatrix} \varepsilon_{rr} & 0 & 0 \\ 0 & \varepsilon_{rr} & 0 \\ 0 & 0 & \varepsilon_{zz} \end{pmatrix}, \quad (2)$$

where the in plane strain components, $\varepsilon_{xx}$ and $\varepsilon_{yy}$, are both equal and labeled here as $\varepsilon_{rr}$, the radial strain component. Each of the off-diagonal shear terms are also zero, and $\varepsilon_{zz}$ represents the axial strain. The strain tensor of Eq. (2) is also constant at all positions throughout the core. Note that, due to the cylindrical symmetry of this structure, the definition of in-plane axes is arbitrary, with Eq. (2) always being of the same form. For use in Eq. (1), the strain tensor of Eq. (2) is then converted into a crystal-oriented coordinate system where the $x'$-, $y'$-, and $z'$- axes are in the [100], [010], and [001] directions, respectively, using the following tensor transformation:

$$\varepsilon'_{ij} = \sum_{pq} a_{ip} a_{jq} \varepsilon_{pq}, \quad (3)$$

where

$$\boldsymbol{a} = \begin{pmatrix} 1/\sqrt{2} & 1/\sqrt{6} & 1/\sqrt{3} \\ -1/\sqrt{2} & 1/\sqrt{6} & 1/\sqrt{3} \\ 0 & -2/\sqrt{6} & 1/\sqrt{3} \end{pmatrix}. \quad (4)$$

In Eq. (3), $\varepsilon'$ is the crystal-oriented strain tensor, $\varepsilon$ is the strain tensor given by Eq. (2), and $\boldsymbol{a}$ is the transformation matrix. This conversion results in the strain tensor:

$$\boldsymbol{\varepsilon'} = \begin{pmatrix} \tfrac{2}{3}\varepsilon_{rr} + \tfrac{1}{3}\varepsilon_{zz} & -\tfrac{1}{3}\varepsilon_{rr} + \tfrac{1}{3}\varepsilon_{zz} \\ -\tfrac{1}{3}\varepsilon_{rr} + \tfrac{1}{3}\varepsilon_{zz} & \tfrac{2}{3}\varepsilon_{rr} + \tfrac{1}{3}\varepsilon_{zz} \\ -\tfrac{1}{3}\varepsilon_{rr} + \tfrac{1}{3}\varepsilon_{zz} & -\tfrac{1}{3}\varepsilon_{rr} + \tfrac{1}{3}\varepsilon_{zz} \\ -\tfrac{1}{3}\varepsilon_{rr} + \tfrac{1}{3}\varepsilon_{zz} \\ -\tfrac{1}{3}\varepsilon_{rr} + \tfrac{1}{3}\varepsilon_{zz} \\ \tfrac{2}{3}\varepsilon_{rr} + \tfrac{1}{3}\varepsilon_{zz} \end{pmatrix}. \quad (5)$$

By substituting the strain tensor of Eq. (5) into Eq. (1) we obtain a non-degenerate singlet mode with $\lambda_s = \lambda_1$ and a degenerate doublet mode with $\lambda_d = \lambda_2 = \lambda_3$:

$$\begin{aligned} \lambda_1 &= \varepsilon_{zz}\left[\tfrac{1}{3}p + \tfrac{2}{3}q + \tfrac{4}{3}r\right] \\ &+ \varepsilon_{rr}\left[\tfrac{2}{3}p + \tfrac{4}{3}q - \tfrac{4}{3}r\right], \\ \lambda_{2,3} &= \varepsilon_{zz}\left[\tfrac{1}{3}p + \tfrac{2}{3}q - \tfrac{2}{3}r\right] \\ &+ \varepsilon_{rr}\left[\tfrac{2}{3}p + \tfrac{4}{3}q + \tfrac{2}{3}r\right]. \end{aligned} \quad (6)$$

As only one Ge-Ge peak was observed in the Raman spectra of strained, core-shell nanowires [Fig. 2], Raman intensity calculations are necessary in order to determine which mode has been measured. Using the polarization selection rule of Eq. (7),

we determine the relative intensity, $I_i$, between the three Ge-Ge modes, $\lambda_i$:[15]

$$I_i \propto \left| \vec{E}_{inc}^T \cdot \boldsymbol{R}_i' \cdot \vec{E}_{scat} \right|^2, \qquad (7)$$

where

$$\boldsymbol{R}_i' = \sum_j \left( \vec{u}_j \cdot \vec{u}_i' \right) \boldsymbol{R}_j. \qquad (8)$$

The unstrained phonon wavevectors, $\vec{u}_j$, are given by $\vec{u}_1 = [100]$, $\vec{u}_2 = [010]$, $\vec{u}_3 = [001]$. $\vec{u}_i'$ are the phonon wavevectors under strain, values used below were found through a solution of the secular equation using a strain tensor in the form of Eq. (2), where $\varepsilon_{rr}$ and $\varepsilon_{zz}$ are arbitrary compressive values. $\vec{E}_{inc}$ and $\vec{E}_{scat}$ are the polarization vectors of the incident and scattered light, respectively. The $\boldsymbol{R}_i'$ in Eqs. (8) and (9) are the Raman tensors of the perturbed system, while $\boldsymbol{R}_j$ are those of the unstrained, cubic system:[33]

$$\boldsymbol{R}_1 = \begin{bmatrix} 0 & 0 & 0 \\ 0 & 0 & d \\ 0 & d & 0 \end{bmatrix}, \quad \boldsymbol{R}_2 = \begin{bmatrix} 0 & 0 & d \\ 0 & 0 & 0 \\ d & 0 & 0 \end{bmatrix},$$

$$\boldsymbol{R}_3 = \begin{bmatrix} 0 & d & 0 \\ d & 0 & 0 \\ 0 & 0 & 0 \end{bmatrix}. \qquad (9)$$

The calculated relative Raman intensities of the singlet ($I_1$) and doublet ($I_2$, $I_3$) Ge-Ge modes using Eqs. (7) - (9) are given in Table I for three separate polarization conditions. We consider only the polarization of $\vec{E}_{inc}$ parallel to the [111] axis due to the enhancement of the parallel electric field magnitude inside of the nanowire, as noted previously. Depending on the polarization direction of the scattered light, three different Raman modes could be active in a given measurement. We note, however, that these calculations do not include the previously discussed "antenna-effect." Analogous to the case of anisotropic incident field to NW coupling, an oscillating, internal electric field will radiate preferentially when its polarization is parallel to the NW axis, as evident in theoretical work[26] and experimental photoluminescence results[34,35]. Therefore, we attribute the sole Raman peak observed experimentally at ~300 cm$^{-1}$ to the singlet mode ($\lambda_1$), for which $\vec{E}_{scat} \parallel \vec{E}_{inc} \parallel [111]$. The singlet peak will have the highest intensity according to Table I, and will also satisfy the polarization memory effect.[26]

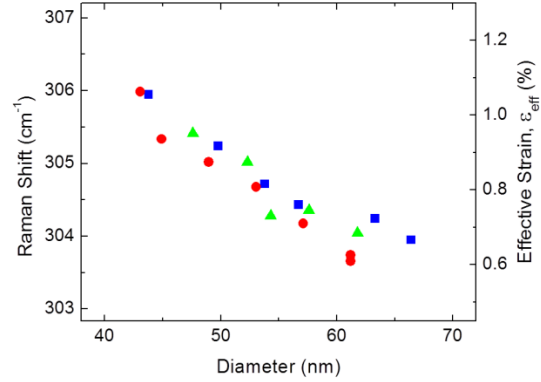

FIG. 3. (Color online) Ge-Ge peak frequency (left axis) vs. diameter, measured in three individual Ge-Si$_{0.5}$Ge$_{0.5}$ core-shell nanowires. The right axis shows the corresponding effective strain defined by Eq. (10).

Table I. Relative Raman intensities of the singlet ($I_1$) and doublet ($I_2$, $I_3$) strained Ge-Ge modes.

| $\vec{E}_{inc}$ | $\vec{E}_{scat}$ | $I_1$ | $I_2$ | $I_3$ |
|---|---|---|---|---|
| [111] | [111] | 12 | 0 | 0 |
| [111] | [1$\bar{1}$0] | 0 | 2 | 0 |
| [111] | [11$\bar{2}$] | 0 | 0 | 6 |

By simplifying the singlet frequency term, $\lambda_1$ in Eq. (6), and substituting the Ge phonon deformation potential values from Table II we obtain a relation between the measured Raman peak frequency and a linear combination of radial and axial strain, which we define as the effective strain:

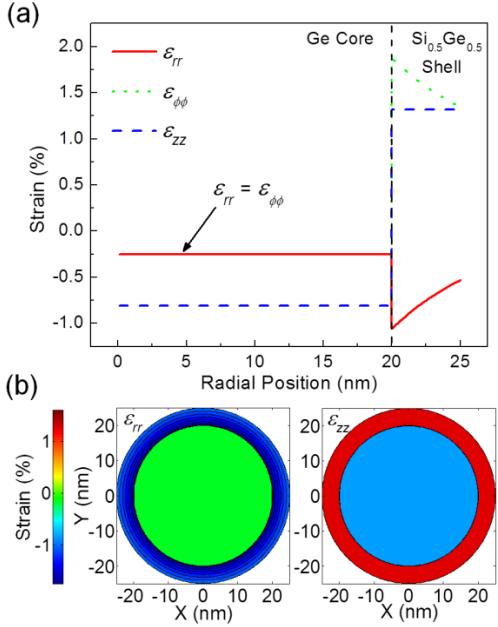

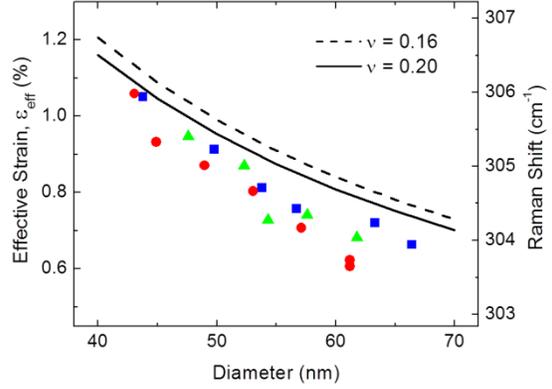

FIG. 5. (Color online) Raman measurement results from Fig. 3, along with calculations of effective strain using a continuum elasticity model (Ref. 13).

FIG. 4. (Color online) (a) Radial cross-section of the three cylindrical strain components ($\varepsilon_{rr}, \varepsilon_{\phi\phi}, \varepsilon_{zz}$) in a Ge-Si$_{0.5}$Ge$_{0.5}$ core-shell nanowire, as calculated from Ref. 13. (b) Contour plots of the radial and axial strain components of the same structure shown in (a). The in-plane Poisson ratio, $\nu_{(111)} = 0.16$, was used here.

$$\varepsilon_{eff} = |\varepsilon_{zz} + 0.729\varepsilon_{rr}|. \qquad (10)$$

Figure 3 data shows the measured Ge-Ge Raman shift (left axis) and the effective strain (right axis) as a function of the diameter at each measurement position. The increase in the Ge-Ge Raman shift with reducing the nanowire diameter can be understood by considering the change in the relative core/shell thickness due to nanowire tapering, discussed above. At large diameter, the strain in this structure can be approximated by that of a planar thin film (the shell) grown on a thick substrate (the core). The shell, therefore, will accommodate nearly the entire lattice mismatch,[36,37] leading to measurements of the core's Ge-Ge peak position nearest to that of the unstrained case, 300.5 cm$^{-1}$. Decreasing the nanowire's total diameter causes an increase in the elastic compliance of the core, allowing it to accommodate a larger total misfit strain. This is evident in the large effective strain measurements of over 1.1 % in core-shell structures near 40 nm in size.

Having established a relationship between the measured Ge-Ge Raman mode frequency and the effective strain, we now compare our measurement results with theoretical calculations of the nanowire strain distribution. Using a continuum elasticity (CE) model, and taking into account surface stress,[13] we calculate the strain distribution in Ge-$Si_{0.5}Ge_{0.5}$ core-shell nanowires. This analytical method relies on the cylindrical symmetry of the nanowire, along with the assumption of isotropic elastic constants, to solve the three-dimensional equations of equilibrium. Since both Ge and Si are crystals with cubic symmetry, care must be taken in choosing the values for Young modulus and Poisson ratio. Table II lists the structural parameters and Ge elastic constants used in the calculation. We use the average $\nu_{(111)} = 0.16$ of the highly anisotropic Poisson ratio within the (111) plane, perpendicular to the nanowire's main axis, combined with the Young modulus value in the [111] crystal direction.[40] We have also performed strain calculations using the Voigt average of the Poisson ratio,[40] $\nu_V = 0.20$. Furthermore, the CE model requires that the elastic constants of the core and shell be identical, an assumption suitable for the Si/Ge system under consideration.

Table II. Values of the normalized phonon deformation potentials, Young modulus ($E$), Poisson ratio ($\nu$), and surface stress ($\tau$) used in the calculations.

| $p/\omega_0^2$ | $q/\omega_0^2$ | $r/\omega_0^2$ | $a_{core}$ (Å) | $a_{shell}$ (Å) | $E_{[111]}$ (GPa) | $\nu$ | $\tau$ (N/m) |
|---|---|---|---|---|---|---|---|
| -1.66[a] | -2.19[a] | -1.11[b] | 5.658[c] | 5.538[c] | 155[d] | 0.16[e], 0.20[f] | 1.00[g] |

[a]Ref. 32.
[b]Ref. 19.
[c]Ref. 38.
[d]Ref. 39.
[e]Average in the (111) plane, Ref. 39.
[f]Voigt average, Ref. 40.
[g]As estimated according to Ref. 13.

Figure 4(a) shows a radial slice of each cylindrical strain component ($\varepsilon_{rr}, \varepsilon_{\phi\phi}, \varepsilon_{zz}$) of a 50 nm diameter nanowire with a shell thickness of 5 nm. The contour plots of $\varepsilon_{rr}$ and $\varepsilon_{zz}$ for the same structure are also shown in Fig. 4(b). As noted previously, each strain component is constant throughout the core, while the magnitudes of $\varepsilon_{rr}$ and $\varepsilon_{\phi\phi}$ are seen to decrease with radial position in the shell. In order to convert the cylindrical strain tensor components of Fig. 4 into a Cartesian coordinate system in the form of Eq. (2), we decouple the radial atomic displacement field, $u_r$, given by Ref. 13, into two separate displacement fields, $u_x$ and $u_y$:

$$u_x = u_r \cos(\theta), \ u_y = u_r \sin(\theta). \quad (11)$$

The Cartesian strain components, $\varepsilon_{xx}$ and $\varepsilon_{yy}$, are then found by differentiation of the appropriate displacement field:

$$\varepsilon_{xx} = \frac{du_x}{dx} = \varepsilon_{rr}, \ \varepsilon_{yy} = \frac{du_y}{dy} = \varepsilon_{rr}. \quad (12)$$

Figure 5 shows a comparison between strain calculations using the CE model (solid and dashed lines), and the experimental Raman data (symbols) of Fig. 3. The experimental and theoretical results follow the same trend as a function of the nanowire diameter, however the degree of quantitative agreement depends on the Poisson ratio. In each case, the measured effective strain is lower than the calculated value, most likely due to strain relaxation through defect formation at the core/shell interface. The measured effective strain values range between 74 and 98% of the calculated effective strain.

## IV. SUMMARY AND CONCLUSIONS

In summary, we measure the Raman spectra of strained, Ge-$Si_{0.5}Ge_{0.5}$ core-shell nanowires with a $Si_{0.5}Ge_{0.5}$ shell thickness of 5 nm. Compared to unstrained GeNWs, the core-shell structures show a diameter-dependent blue-shift of the core's Ge-Ge vibrational mode, with the shift increasing with reducing the nanowire diameter. We use lattice dynamical theory to convert the Raman shift results into an effective strain: $\varepsilon_{eff} = |\varepsilon_{zz} + 0.729\varepsilon_{rr}|$. The effective strain determined from Raman spectroscopy is in good agreement with continuum elasticity calculations of the strain distribution in our Ge-$Si_{0.5}Ge_{0.5}$ core-shell nanowires.

## ACKNOWLEDGEMENTS

This work has been supported by NSF grants DMR-0846573 and DGE-0549417, along with the Texas State ARP-ATP program.